# A Novel Smart Memory Alloy Recentering Damper for Passive Protection of Structures Subjected to Seismic Excitations Using High Performance NiTiHfPd Material

Farzad Shafiei Dizaji[1] and Mehrdad Shafiei Dizaji[2]

**Abstract**

This research proposes and evaluates a superelastic memory alloy re-centering damper system for improving the reaction of steel frame buildings that have been exposed to several levels of seismic threat. The planned superelastic memory alloy re-centering damper (SMARD) relies on high-performance shape memory alloy (SMA) bars for its abilities of recentering and augments its deformation potential with friction springs. To begin, this study investigates the superelastic reaction of NiTiHfPd SMAs under a variety of conditions and shows how they can be used in seismic applications. To gather experimental results, uniaxial experiments on superelastic NiTiHfPd SMAs are performed at temperatures ranging from -35 to 25 oC and loading frequencies ranging from 0.05 to 1 Hz with four distinct strain amplitudes. We explore the impact of loading rate and temperature on the superelastic properties of NiTiHfPd SMAs. The complex answer of 6-floor and 9-floor steel special moment frame buildings with built SMARDs is then determined using an empirical model. Finally, nonlinear reaction time background simulations are used to characterize the actions of managed and unregulated buildings while 44 ground motion data are used. The results indicate that SMARDs will significantly reduce the dynamic behavior of steel-frame buildings at various seismic threat levels while simultaneously improving their post-earthquake functionality.

## 1. INTRODUCTION

Traditional seismic design proposes prioritize the capacity of buildings to subside seismic loads through plastic deformations in engineered areas of the steel frames, suggesting significant structural collapse and therefore the possibility of residual drifts following a strong earthquake [1]. Peak reaction measures, such as story drifts and floor accelerations, are often used to determine the seismic performance of different structural systems. However, numerous experiments have demonstrated that residual drifts caused by the nonlinear behaviour of yielding components in a structural framework may be critical in defining a structure's output following a seismic event and in assessing possible harm [2-3].

---

1. Researcher and Adjunct Professor, Department of Engineering & Society, University of Virginia, Charlottesville, VA 22901 (Email: ffs5da@virginia.edu)
2. Ph.D. Former Researcher at the Department of Civil Engineering and Environment, University of Virginia, Charlottesville, VA 22901 (Email: ms4qg@virginia.edu)

McCormick et al. [4] investigated the impact of residual drifts on inhabitants. They found that residual drifts greater than 0.5 percent in structures can indicate the structures complete failure from an economic standpoint. Erochko et al. [5] conducted another analysis in which they conducted the residual drift reaction of unique moment resisting frames (SMRFs) and also buckling restrained braced frames (BRBFs). Both forms of building structures exhibit substantial residual drift, with values ranging between 0.8–1.5 percent for SMRFs and 0.8–2.0 percent for BRBFs when subjected to design-based excitations. Ramirez and Miranda [6] discovered that taking residual drift into account when estimating building earthquake damages greatly raises the estimated economic losses. By minimizing residual drifts in a building that has been exposed to a seismic incident, structural engineers may optimize post-event functionality, lower maintenance costs, and improve public protection.

To improve the seismic stability of structural systems, it is desirable to provide structural systems capable of safe energy dissipation and complete self-centering. These self-centering or recentering structures display a flag-shaped hysteric reaction and the potential to revert to either little or no deformation after each loop. The self-centering Feature enables for the control of structural damage while minimizing residual drift. Numerous self-centering solutions have been developed during the last two decades, namely, rocking wall panels, post-tensioned mechanisms, and self-centering support platforms [7–12]. [13] Examines self-centering systems in detail. Shape memory alloys (SMAs) are a variety of intelligent materials that manifesto remarkable properties such as exceptional re-centering capacity, greater corrosion and fatigue resistance, and a high energy attenuation capability. SMA's peculiar properties have piqued researchers' interest in designing seismic control schemes focused on SMA [14]. Numerous studies on the use of SMAs in the creation of an effective energy dissipation device with self-centering capability have been performed [15–20].

Numerous passive, active, and semi-active control systems were suggested and designed to reduce the negative impacts of complex environmental threats such as earthquakes, heavy winds, and hurricanes and to establish a more resilient configuration under dynamic loads [21]. Passive devices have become the most extensively tested and generally deployed strategies because they are efficient, do not need external control, and just never bring down the system. However, integrating passive structures into buildings that increase efficiency over a broad spectrum of earthquake and wind vibration is a difficult challenge.

Numerous passive energy dissipation devices were designed to help structures withstand the destructive impact of natural hazards [22]. Passive energy dissipation systems fall into two broad categories: hysteretic and rate-dependent. Metallic yielding devices and friction devices are examples of hysteretic machines. In hysteretic applications, energy dissipation is mainly determined by relative displacements inside the system. These instruments have hysteretic energy dissipation by initially stiffening the system before yielding or slip occurs. They do not, however, have enough damping for minor vibrations induced by wind excitation or repeated seismic activities and, as a result, raise the forces and accelerations acting on the system due to their strong stiffness. Additionally, a metallic device typically has a finite amount of operating cycles and may need replacement after a severe accident, whereas friction devices may result in irreversible deformations in the absence of a restoring force mechanism. Viscoelastic dampers and Fluid viscous dampers are examples of rate-dependent systems. The ability of these devices to dissipate energy is dependent on the velocity around the system [22]. Viscoelastic dampers have a low force and damping power, while viscous dampers have a strong force and damping capacity. In general, rate-based devices can decrease substantially throughout all motion magnitudes but lack the energy dissipation power of hysteretic devices [23]. Notably, these devices lack a self-centering feature that would enable the structure to revert to its original location.

Numerous attempts have been made to build novel structural structures capable of stable energy dissipation and complete self-centering [24]. Numerous researchers have suggested form memory alloy (SMA)-based seismic safety systems, including but not restricted to SMA-based bracing systems [25-27]; SMA-based dampers [28-31]; and SMA-based isolation systems [32-36]. Among the different SMA compositions tested, the NiTi alloy has received the most attention for use in SMA-based dampers. Due to the limited energy dissipation ability of NiTi SMAs, damping in certain devices has been augmented by pre-tensioning SMA components [37], integrating frictional devices [38-40], energy-absorbing steel struts [41], viscoelastic devices [42-43], or buckling-restrained braces [44].

A newly created class of SMAs composed of NiTiHfPd exhibits greater energy dissipation and high tension [45] and thus overcomes the shortcoming of currently available SMAs. At room temperature, Figure 1 shows standard strain-stress diagrams for NiTi and NiTiHfPd SMAs. As can be shown, the region contained inside the hysteresis ring, which represents the dissipated electricity, is significantly greater for NiTiHfPd than for NiTi. NiTiHfPd still has a considerably higher strength than NiTi. Thus, the cross-sectional region of the SMA elements used to develop the damping device's design force would be significantly smaller for NiTiHfPd. Due to the strong energy dissipation ability and power of NiTiHfPd, a vibration control system built on it would eliminate the need for external energy dissipation units and/or significant quantities of content. This study explores the experimental characterization of NiTiHfPd SMAs and the computational design of a basic yet effective energy dissipating and re-centering system based on NiTiHfPd SMAs that eliminates the need for additional fabrication and material costs.

This article suggests a computational design for an advanced SMA-based damper and then numerically simulates its performance using a case study. The superelastic memory alloy re-centering damper system (SMARD) blends the energy-dissipating capability of a cold friction spring with the superior re-centering features of SMA bars in a unified device, resulting in an enhanced hybrid reaction. The following sections explain the experimental studies conducted on the hybrid damper's individual parts, including SMA bars and a cold friction spring. Then, the theoretical superelastic memory alloy re-centering damper device's architecture and action was discussed. Following that, the modeling of SMARD is debated, as well as the modeling of a 6-story and a 9-story steel frame structure. Finally, nonlinear reaction background simulations are used to determine the efficacy of the SMARD in controlling the steel building's response.

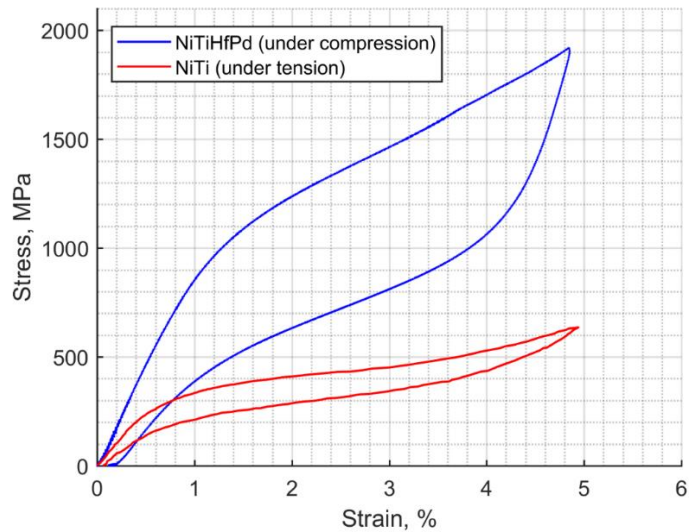

**Figure 1.** Graph of stress-strain for NiTiHfPd and NiTi at room temperature and loading frequency of 1.0 Hz

## 2. EXCELLENT DAMPING AND STRENGTH SMAs made of NiTiHfPd

Due to its excellent shape memory properties, corrosion tolerance, biocompatibility, and ductility, NiTi SMAs were researched and used in a number of applications. They do, however, have certain shortcomings that hinder their usefulness in some applications. Alloying is a highly efficient method for engineering properties and resolving these issues. For example, in order to be used as dampers, an SMA material must exhibit a high degree of mechanical hysteresis and transforming strain at high stress levels.

Although the mechanical hysteresis and damping power of binary NiTi alloys are 200-400 MPa and 16 J.cm-3, respectively [47], they can indeed be boosted to 500-600 MPa and 38 J.cm-3 with the inclusion of Nb [48]. SMAs have also been studied with the incorporation of elements such as Zr, Hf, and Pd to increase their efficiency, wear resistance, and damping capability. Due to its low expense, moderate ductility, and increased work performance, it was determined that the integration of Hf to NiTi binaries was promising [49-50]. However, prior NiTiHf alloys could be used in realistic implementations; one of the primary difficulties that must be overcome is their brittleness. As a result, the literature has investigated the addition of a number of quaternary components to NiTiHf. Cu addition to NiTiHf, for example, has been shown to decrease thermal hysteresis, increase thermal stability, ductility, and two-way shape memory effect, but has had no discernible effect on the alloy's other shape memory properties [51-52]. The inclusion of Pd as a quaternary feature to NiTiHf alloys significantly improves ductility, TT change, and shape memory behavior. Pd may be applied in place of Ti to raise transition temperatures, or in place of Ni to boost superelasticity. A single crystalline NITiHfPd alloy has been stated to exhibit enormous mechanical hysteresis of up to 1270MPa, superelasticity at incredibly high-stress levels (2.5GPa), and a damping capability of 44 J cm3 [46, 52]. Even polycrystalline NiTiHfPd alloys have strong work outputs of 32–35 J cm-3 (up to 120 oC), which are significantly higher than those of NiTi and Ni. NiTiHfPd SMAs are an excellent candidate for applications requiring high damping due to their ultra-high power, outstanding damping capability, and realistic capabilities to display super-elastic activity at room temperature. Apart from alloying, the most frequently used techniques for enhancing the shape memory and mechanical properties of SMAs are thermomechanical treatment (e.g. cold working and post-annealing), precipitate forming, and grain refining of polycrystalline alloys. Thermal therapies, on the other hand, seem to be the most realistic and cost-effective procedure. It has been shown that the development of precipitates increases the matrix's strength and necessary essential tension, resulting in improved shape memory reaction, fatigue resistance, and cyclic stability. It should be remembered that the precipitates' capacity to improve is dependent on their duration, volume fraction, antiparticle spacing, and coherency [53].

3. **EXPERIMENTAL TESTING OF NiTiHfPd SMAs**

For experimental characterization, induction melted $Ni_{45.3}Ti_{29.7}Hf_{20}Pd_5$ (at %) alloys are used. The sample dimensions are 4x4x8mm3. The sample is heat-treated at 400°C for 3 hours, and then water quenched before testing. Mechanical tests are performed using a 100 kN MTS Landmark servo-hydraulic test frame. Loading and unloading are conducted at the selected frequencies of 0.05, 0.5, and 1Hz and temperatures of -35°C, 5°C, and 25°C. The strain is determined by an MTS high-temperature extensometer with a gauge length of 12 mm, and K-type thermocouples are used to monitor the temperature during the test.

4. **EXPERIMENTAL RESULTS**

Fig. 2(a) to (c) show strain-stress curves of NiTiHfPd SMAs at -35 °C, 5 °C, and 25 °C, respectively. The curves at each temperature are plotted for the tests conducted at different strain amplitudes and different loading frequencies. As can be shown, the hysteresis loops of NiTiHfPd SMAs move upward with increasing temperature. In addition, when the temperature decreases from 25°C to -35°C, the hysteresis loops tend to have a wider area. This wider area indicates that the energy dissipation of the SMA increases with a decrease in temperature. Increasing loading frequency also shifts hysteresis loops upward, yet, it does not change the area of hysteresis loops considerably [54].

To make a more quantitative evaluation of the testing results, some significant mechanical properties, including the equivalent (secant) stiffness, $K_e$, the energy dissipation per cycle, $E_D$, and the equivalent viscous damping ratio, $\xi_{eqv}$ are calculated. The equivalent viscous damping ratio and the equivalent (secant) stiffness are defined as:

$$\xi_{eqv} = \frac{E_D}{4\pi E_S} \quad (1)$$

$$K_e = \frac{F_{max}-F_{min}}{D_{max}-D_{min}} \quad (2)$$

Where $E_S$ denotes the maximum elastic strain energy per loop; and $F_{max}$ and $F_{min}$ denote the maximum and minimum forces produced for the maximum and minimum cyclic deformation $D_{max}$ and $D_{min}$.

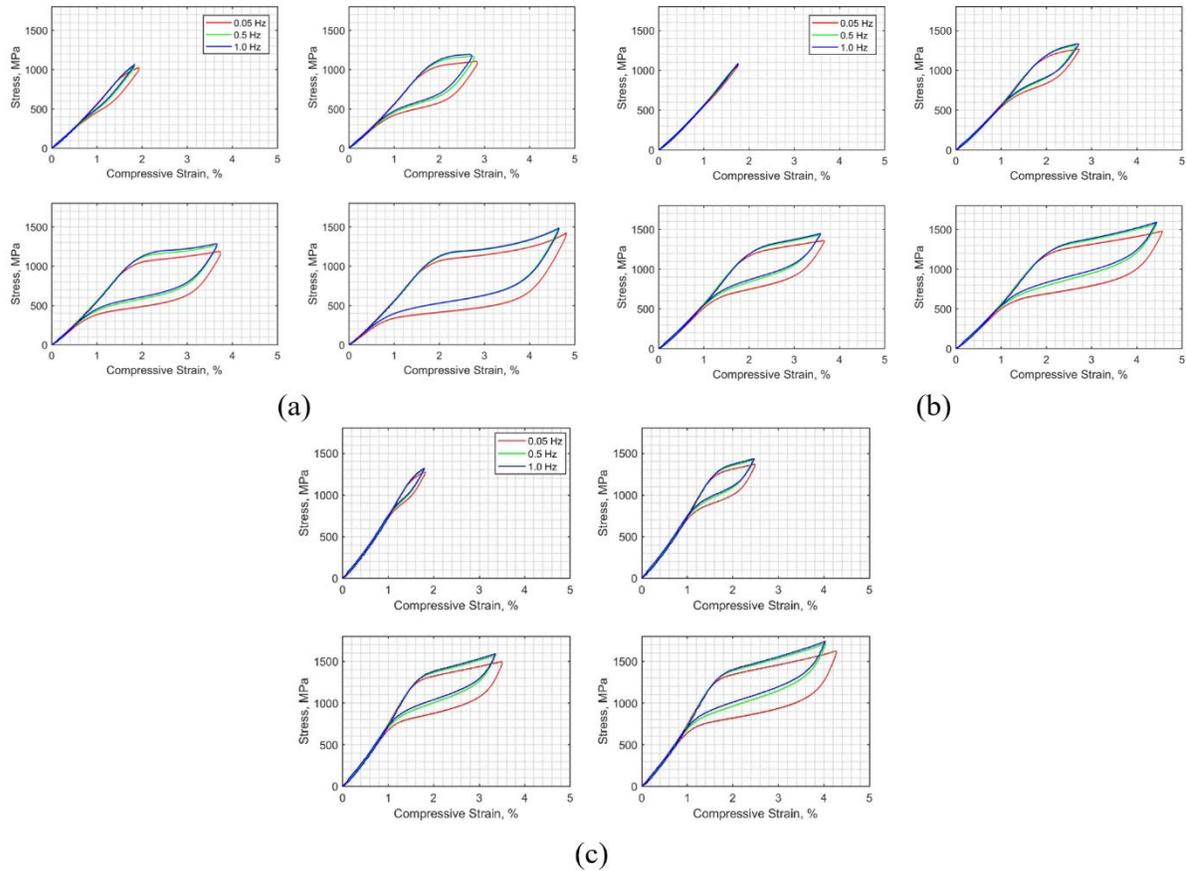

**Figure 2.** Stress-strain curves of SMAs at different temperatures; (a) -35°C, (b) 5°C, and (c) 25°C with various strain and loading frequency levels

Fig. 3 shows the variation of energy dissipation per cycle, equivalent viscous damping, and equivalent stiffness with temperature for the tests conducted at different loading frequencies. It can be seen that the energy dissipation generally decreases with increasing temperature. In particular, there is a 32%, 42%, and 50% decrease in the dissipated energy at the loading frequencies of 0.05, 0.5 and 1.0 Hz, respectively when the temperature changes from -35 °C to 25 °C. A similar decrease is also observed in the equivalent viscous damping ratio for increasing temperature. The equivalent viscous damping ratio decreases from 4.96% at -35 °C to 3.34% at 25 °C at a loading frequency of 0.05 Hz; from 4.04% at -35 °C to 2.29% at 25 °C at a loading frequency of 0.5 Hz; and from 4.07% at -35 °C to 1.98% at 25 °C at a loading frequency of 1 Hz. Despite the significant changes in damping properties observed for a temperature change from -35 °C to 25 °C, the difference in both dissipated energy and equivalent viscous damping ratio remains within 10% when the temperature changes from 5 °C to 25 °C. On the other hand, equivalent stiffness obtains higher values at higher temperatures for various loading frequencies. When temperature increases from -35 °C to 25 °C, the equivalent stiffness increases by 29%, 34%, and 35% at loading frequencies of 0.05, 0.5, 1 Hz, respectively. This increase in the equivalent stiffness is due to the reality that during loading, the material reaches higher stress levels for the same strain amplitude at higher temperatures.

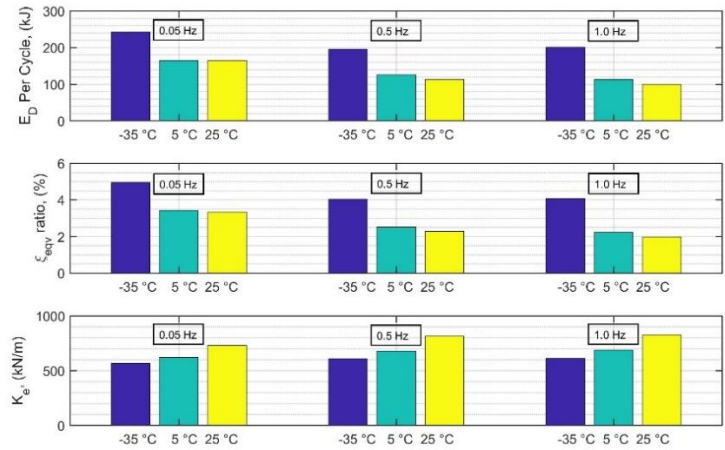

**Figure 3.** Variation of energy dissipation per cycle, equivalent viscous damping, and equivalent stiffness with trempreture

Figure 4 shows the variation of energy dissipation per cycle, equivalent viscous damping, and equivalent stiffness with loading frequencies at different temperatures. It can be seen that the reduction in energy dissipation for various temperatures ranges from 20% to 31% as the loading frequency increase from 0.05 Hz to 0.5 Hz. Nonetheless, the energy dissipation alters only -10% when loading frequencies change from 0.5 Hz to 1.0 Hz. Similarly, it can be seen that the equivalent viscous damping ratio decreases by increasing loading frequency. Specifically, there is a 26% and 11% reduction in the equal viscous damping ratio when the loading frequency changes from 0.05 Hz to 0.5 Hz and when the loading frequency alters from 0.5 Hz to 1.0 Hz, respectively. On the other hand, due to an increase observed in the slope of forwarding phase transformation plateau with an increase in the loading frequency, the equivalent stiffness attains slightly higher values at more significant loading frequencies. As the loading frequency increases from 0.05 Hz to 1.0 Hz, the equivalent stiffness rises 8%, 10%, and 13% at test temperatures of 35 °C, 5 °C to 25 °C, respectively.

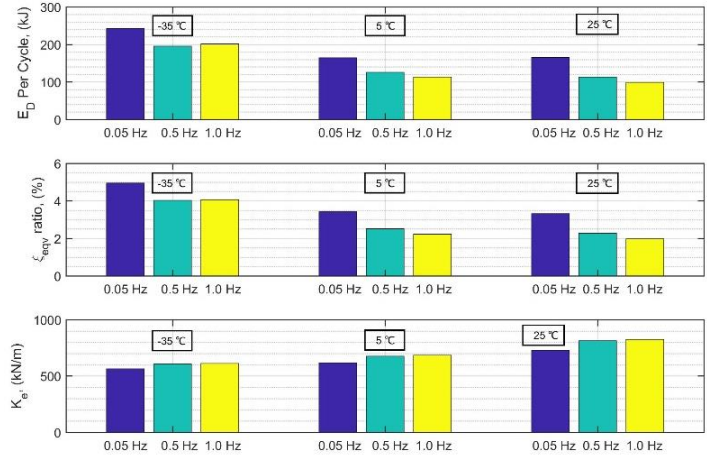

**Figure 4.** Variation of energy dissipation per cycle, equivalent viscous damping, and equivalent stiffness with loading frequency

## 5. HIGH DAMPED COLD FRICTION SPRING

Friction springs are made up of precision-machined outer and inner circles that strike on their tapered sides. This contact surface is referred to as the feature. Figure 5 illustrates a friction spring made up of four outer circles, three inner rings, and two inner ring halves, totaling eight components. The amount of elements is critical for spring design since it dictates the complete travel and function of the spring. Friction springs do well when they terminate in inner ring halves. As a result, an even number of components is preferred. As the spring shaft is filled axially, the tapered surfaces converge, expanding the outer rings and contracting the inner rings. 2/3 of the energy applied would be consumed by the tension between the mating rings' tapered surfaces.

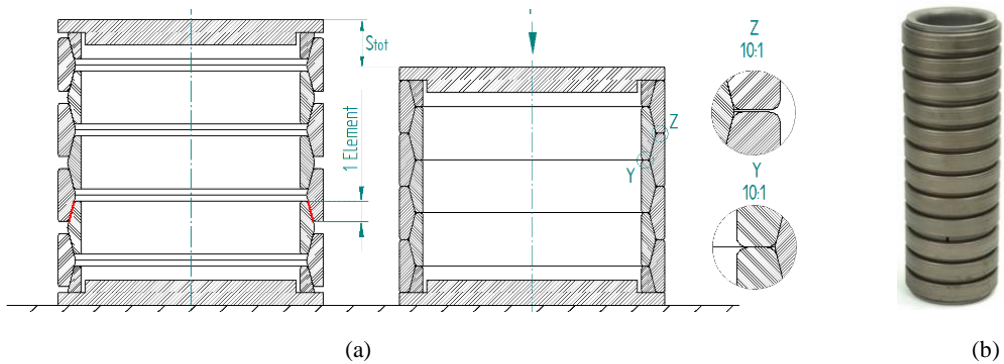

(a)                                                            (b)

**Figure 5.** (a) Configuration of a friction spring with 8 elements, at rest (left) and loaded (right), (b) friction spring

Figure 6 shows the basic technical characteristics of friction spring. The ring size collection describes additional critical friction spring parameters: the end force F, the outer and inner lead diameters D2 and d2, the element height (he), the spring trip per element (se), and the energy absorption per element We. The requisite number of elements e is determined for a specific application by subtracting the appropriate spring travel from the necessary spring job W, thus ignoring the pre-load. Finally, the independent rings are connected together to create full spring columns.

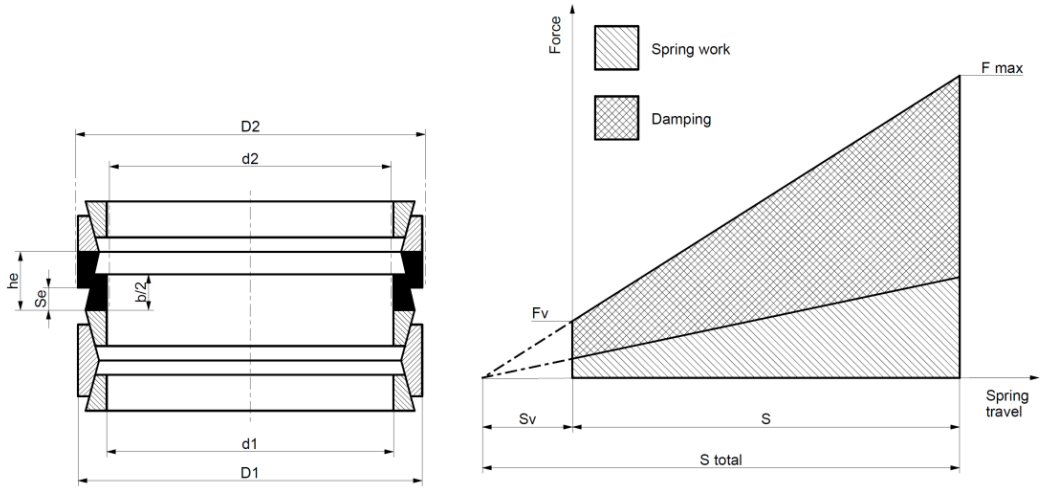

**Figure 6.** Basic technical characteristics of friction spring

Figure 7 shows experimental test results of the sample friction spring loaded up to 15 kN and then unloaded. As it can be seen from the results, friction spring has high energy dissipation capacity and capable of returning to its initial position, which gives the sample excellent re-centering behavior. Moreover, the behavior of the numerical model of the spring at large displacement levels with the experimental behavior of the spring at large displacement levels (i.e., the blue line in figure 8 (a-b) below) is compared in Figure 8 (c). Our numerical model seems to model the spring behavior correctly. Since the device will have a force capacity of 3600 kN, four friction springs, with a force capacity of 900 kN each, will be assembled in parallel with each other in the middle part of the device. The mechanical characteristics of the spring can be obtained accordingly.

Our spring consists of 12 elements type 24301. The deformation capacity of the single spring will be: 12.0 x 8.0 ($S_e$) = 96.0 mm, spring work (absorbed energy) will be: $W_e$ = 3708.0 x 12 = 44496 J, spring length will be: 12 x 33.3 ($h_e$) = 400 mm. By adding additional elements, the stroke (length of the spring), the absorbed energy (spring work) will be increased consequently, but the end force will remain the same, which is 900 kN for selected friction spring.

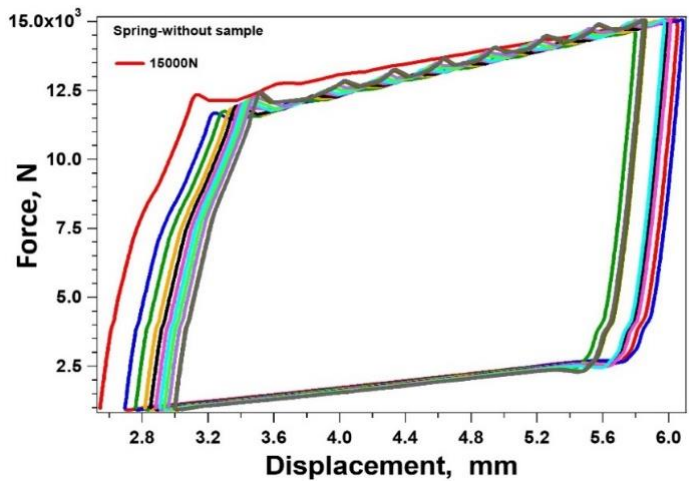

**Figure 7.** Compression behaviour of spring under 15KN

## 6. SHAPE MEMORY ALLOY- BASED RE-CENTERING DAMPER

This paper proposes a modern passive vibration system based on the beneficial properties of NiTiHfPd SMAs. The system is referred to as the SMARD (Shape Memory Alloy-based Re-centering Damper). Figure 9 depicts the SMARD in its initial and displaced roles to illustrate its operating theory. Figure 10 also has a three-dimensional representation of the unit. The SMARD is composed of an SMA-Spring assembly (in yellow), a piston (in gray), and a tightly assembled baffle plate enclosure (in black). Between SMA categories, the large compressive steel springs ensure significant deformations underneath load while still transmitting energy across SMA groups. Voids in the baffle plates' centers enable the piston to freely move while restraining the otherwise free-moving SMA-Spring assembly. Notably, NiTiHfPd SMAs show a more superelastic reaction to compression than NiTiHfPd SMAs. As a result, regardless of whether the piston is dragged out or pushed into the machine, the SMA-Spring assembly can compress. Due to the intrinsic superelastic nature of NiTiHfPd SMAs, the SMARD has an outstanding re-centering capability and a high potential for energy dissipation [55].

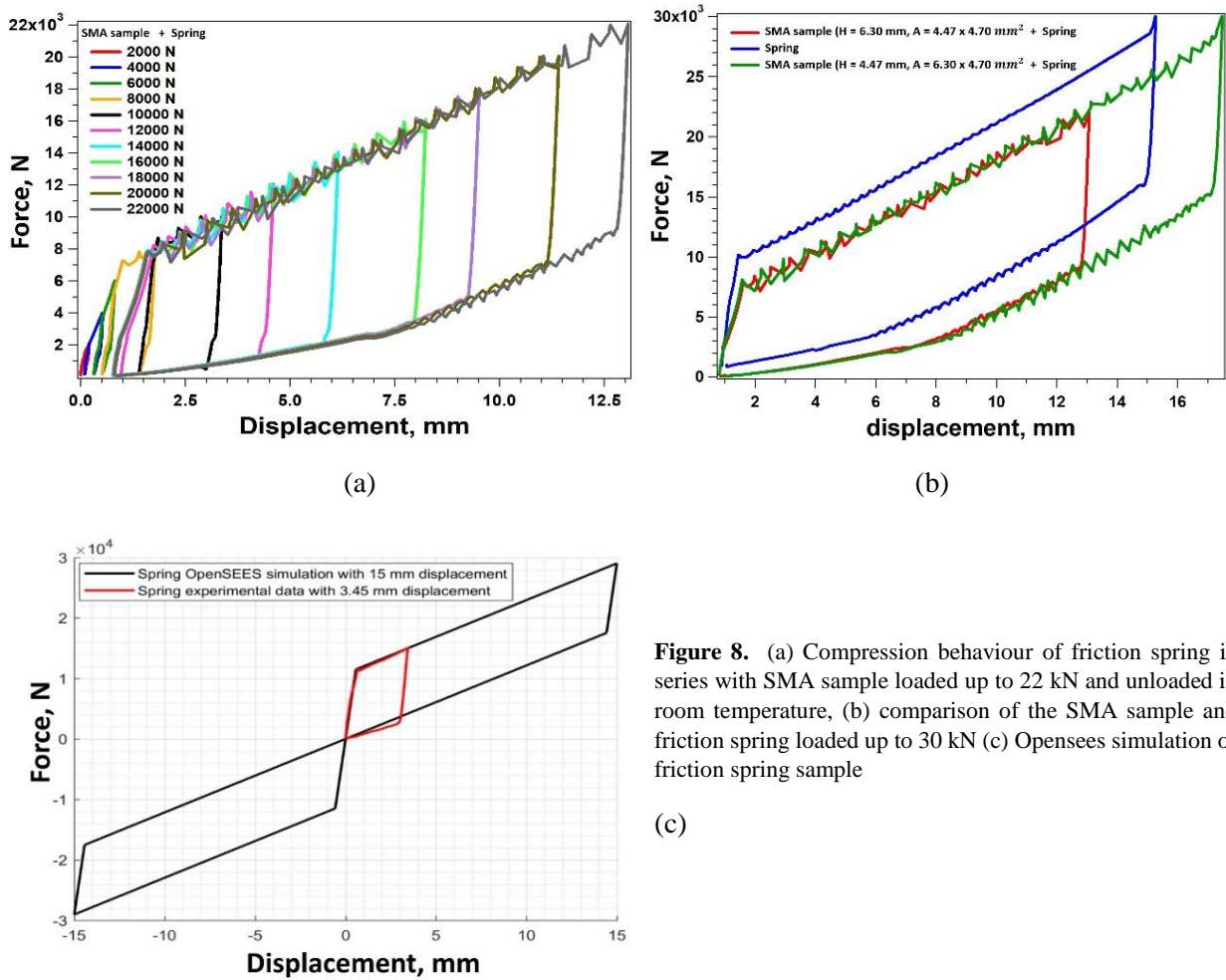

**Figure 8.** (a) Compression behaviour of friction spring in series with SMA sample loaded up to 22 kN and unloaded in room temperature, (b) comparison of the SMA sample and friction spring loaded up to 30 kN (c) Opensees simulation of friction spring sample

Additionally, the system may integrate SMA bars of up to four separate diameters. This integration enables the system to generate a damping force necessary to dissipate energy at varying degrees of excitation. The SMARD is shown in Figure 9 using two separate size bars (SMA1 and SMA2). An adequately built SMARD would demonstrate re-centering and energy-absorbing actions at all modes of vibration within both earthquake and wind oscillations due to the arrangement of SMA bars in the system and the intrinsic re-centering behavior of SMAs. During a mild to moderate case, the SMA bars with details suggest will endure considerably high compressive strains and dissipate the steam.

In contrast, the stiffer SMA bars will experience relatively small compressive strains. During a mild to severe case, all SMA groups participate and dissipate significant quantities of energy, whilst the system provides a greater damping power. The SMARD can stiffen during an extreme case owing to the secondary hardening of SMA bars against massive strains. This extra stiffening would protect the structure from possibly disastrous structural displacements and in the event of unexpectedly powerful ground motions.

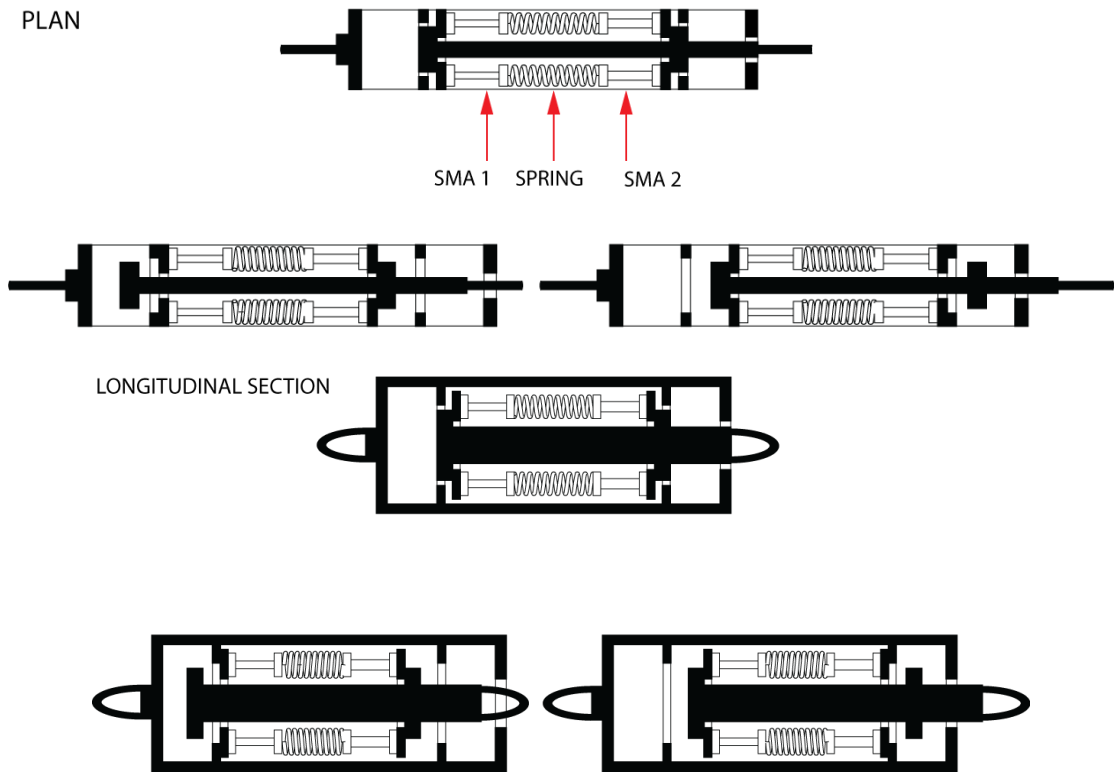

**Figure 9**. Plan view of SMARD at its original and displaced positions

Compared to traditional structural control systems, the proposed SMARD is an attractive alternative in both performance and design efficiency. Due to the high strength of NiTiHfPd SMAs, a compact SMARD device with large force capacities can be efficiently designed by adjusting the total area of SMA bars used in the machine. Here, the prototype SMARD will be designed to have a maximum force capacity of 3600 kN, a stroke of 150 mm, and a loss factor of over 0.30. The loss factor, which is associated with energy dissipation of the device, is computed as dissipated energy per cycle (the area inside the hysteresis loop) divided by the product of $2\pi$ and maximum strain energy (the area under a complete cycle). The target design force can be achieved using four 25-mm diameter and four 25-mm diameter NiTiHfPd SMA bars in a single device, assuming the developed alloys will show superelastic response up to a stress level of 1800 MPa and two different diameter SMA bars are employed. It should be noted that a SMARD with different diameters and lengths of SMA bars could be used to modify the damping, force, and stroke capacities of the device. Thus, it provides an ability to tailor the properties of SMARD for specific applications.

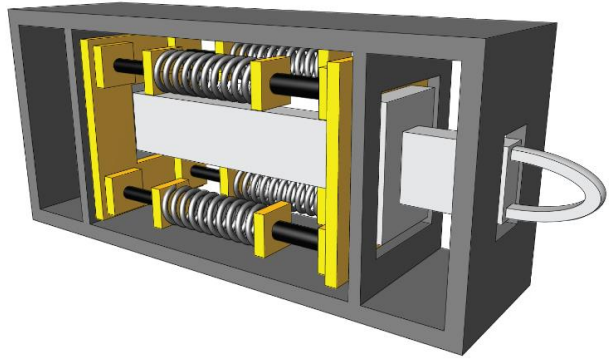

**Figure 10.** Three-dimensional rendering of SMARD

The SMARD is built for a force capacity of 3600 kN and a displacement capacity of 150 mm in this use. To accomplish these layout goals, the SMA bar's cross-sectional region is set to:

$$A_{SingleSMA} = \frac{F}{\sigma} = \frac{900000}{1800} \approx 500 \ mm^2$$

Then SMA bar with the diameter of $D = 25 \ mm$ is selected. The length of the SMA bars should be designed as well. Since all parts of the SMARD device, 2 groups of SMAs plus one group of friction springs, will be under compression, the possibility of the buckling for SMA bars should be considered in the design. The length of the SMA bar is selected to be $L = 200 \ mm$ in order to avoid buckling.

$$P_{cr} = \frac{\pi^2 \cdot E \cdot I}{(k \cdot L)^2}$$

Where $P_{cr}$ stands for buckling force, $E$ is the module of elasticity for the SMA bar, $I$ is a moment of inertia for the circle area of the SMA bar, $k \cdot L$ is the effective length of the SMA sample bar.

$$P_{cr} = \frac{\pi^2 \times (78000) \times 19165}{(0.5 \times 200)^2} = 1473 \ kN > 900 \ kN$$

Therefore, the cross-sectional region chosen corresponds to two groups of four collection SMA bars, each with a diameter of 25 mm. To achieve optimum stroke power, the friction spring must be 400 mm in length.

## 7. STEEL MOMENT FRAMES WITH INSTALLED SMARDs

### 7.1. Buildings description

#### 7.1.1. A six-story building

To assess the proposed damper's effectiveness at minimizing seismic reaction of buildings, 2 special steel moment resisting frames are chosen for this purpose. For numerical analysis, a six-story steel building outlined in FEMA P-751, NEHRP Suggested Seismic Regulations: Design Examples [56] is chosen. The structure is intended to serve as just an office complex in Seattle, Washington, on class C soil. The lateral load resistance of the construction is provided by the unique steel moment-resisting frameworks along the circumference of the house. The structure is composed of five bays measuring 8.53 meters (28 feet) north-south (N–S) and six bays measuring 9.14 meters (30 feet) east-west (E–W). Figure 11 depicts the building's plan and height in the north-south direction. This thesis conducts all studies for lateral loads working in the N–S course. Every floor has a story height of 3.81 m (12 feet-6 in.) except the first, which has a story height of 4.57 m. (15 feet). The analysis is from one of the perimeter steel unique moment frames that act as the structure's seismic-force resisting mechanism. In the north-south direction, both columns curve along their strong axes, and the girders are connected using entirely welded moment-resisting ties. The building is constructed using Reduced Beam Section (RBS) connections in compliance with ASCE/SEI 7-05 [57] and ANSI/AISC 341-05 [58] architecture specifications. The structure is classified as having a Seismic Design Classification (SDC) of D.

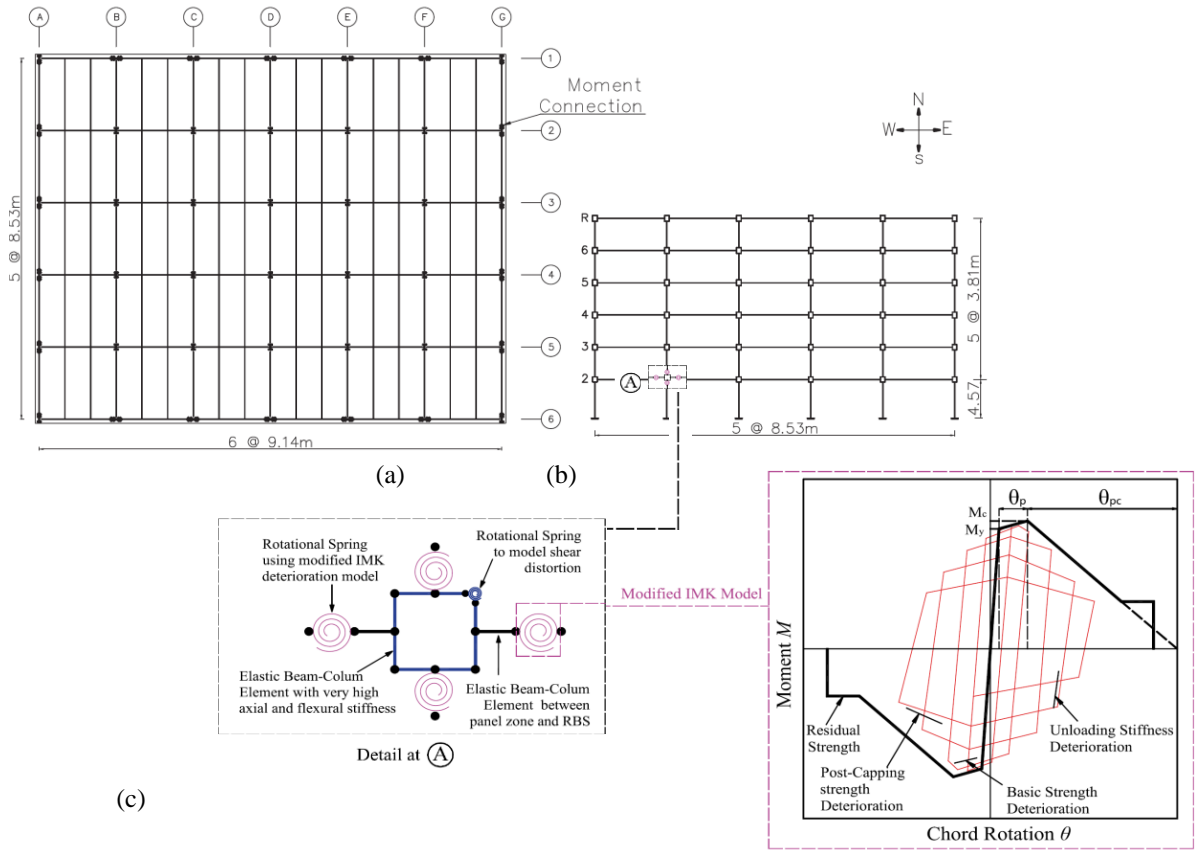

**Figure 11.** 6-story steel special moment-resisting frame: (a) Plan (b) elevation of 6-story steel special moment-resisting frame (c) modeling details.

SDS = 0.912 g and SD1 = 0.459 g are the design spectral acceleration parameters, while SMS = 1.368 g and SM1 = 0.689 g are the highest considered spectral acceleration values. The second level's seismic mass is 1.248 x 106 kg, the third and sixth levels' seismic mass is 1.242 x 106 kg, and the roof level's seismic mass is 1.237 x 106 kg. The structure's fundamental time is 2.04 s. The structure was studied in FEMA P-751 using a nonlinear answer background study, and it was determined that there were excessive story drifts, especially in the lower floor levels. A damping device was proposed to improve the building's efficiency. Dampers made of superelastic material will be used in this study to improve the seismic performance of the chosen structure.

7.1.2. A nine-story building

Additionally, a nine-story steel structure [59] from a SAC steel project is chosen pertaining to computational analysis. The chosen structure has a basement floor and nine floors above ground and was built as an office building in Seattle, Washington, on a stiff soil site (Site Class D). Figure 12 depicts the nine-story building's floor plan and height. The structure is divided into five bays measuring 9.15 m (30 ft) from each direction. Each floor has a story height of 3.96 m (13 ft) except the first, which has a story height of 5.49 m (18 ft), and the basement, which has a story height of 3.65 m. (12 ft). In both directions, the lateral load resisting mechanism is comprised of two special moment resisting frames located along the circumference of the house. The aim of this thesis is to examine one of the moment frames in the E-W path. The seismic masses for floor levels 2–9 are 1.01 106 kg, 9.89 105 kg, and 1.07 106 kg, respectively. At the base of each column, it is presumed that they are pinned. Additionally, the exterior columns on the ground floor are laterally restrained. The building is built as a Risk Category II structure in accordance with ASCE 7-10 [60] utilizing a nonlinear reaction background technique. In the following application reaction, the location is given spectral values. based on the Seismic Design Category (SDC) D: SDS = 0.912 g and SD1 = 0.530 g for the model-specific earthquake (DBE), and SMS = 1. 368 g and SM1 = 0.795 g for the highest considered earthquake (MCE).

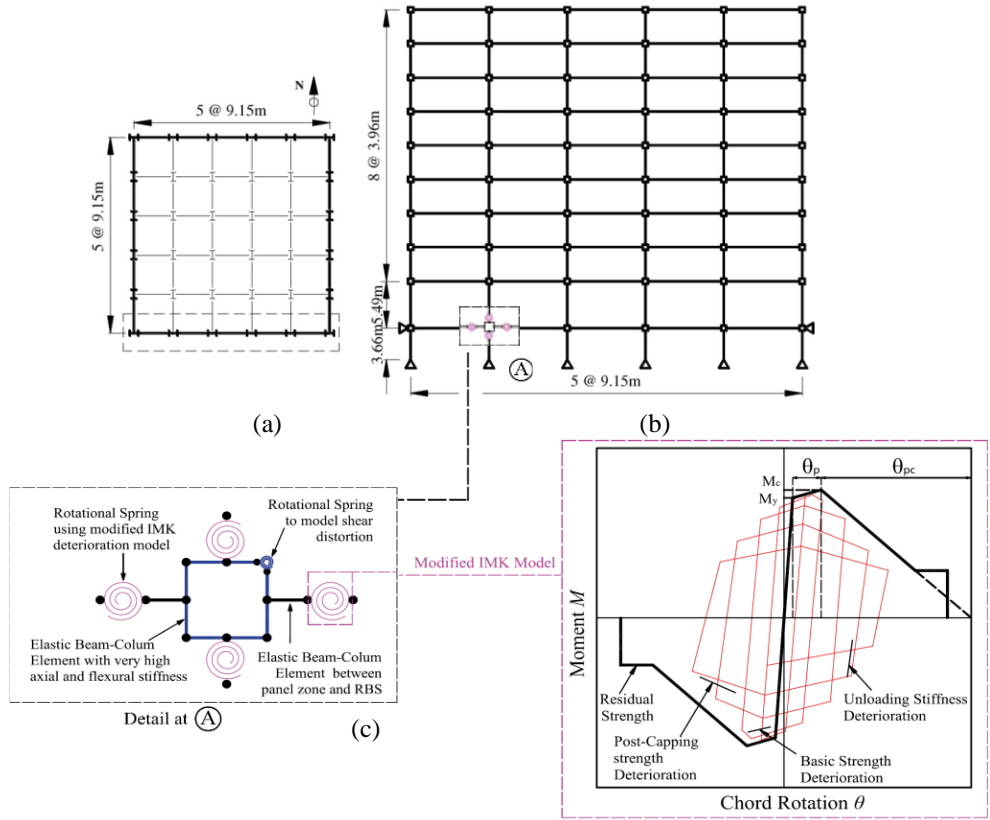

**Figure 12.** 9-story steel special moment-resisting frame: (a) Plan (b) elevation of 9-story steel special moment-resisting frame (c) modeling details.

The two steel frame buildings were numerically modeled in two dimensions using the nonlinear seismic analysis software OpenSees [61]. The beam and column elements are designed using the adapted Ibarra-Krawinkler deterioration model [62] using elastic beam-column elements joined via zero-length inelastic plastic hinges. The updated Ibarra-Krawinkler model of degradation takes bilinear hysteric reaction activity into account. The model parameters for cyclic degradation of zero-length rotational springs were allocated based on those defined by Lignos and Kwawinkler [63]. The yield resistance, unloading stiffness, post-capping strength, and reloading stiffness of rotational springs all signify their deterioration characteristics. The elastic stiffness, plastic rotation, a post-capping plastic rotation power, and the resulting residual strength all contribute to the moment-rotation curve. It is believed the structural steel has a yield stress of 375 MPa. To seize the significant panel zone deformation types, the panel zones are developed using the Krawinkler model [64]. Four rigid ties and a rotational spring in the upper right corner of the Krawinkler model reflect shear distortion in the panel field. Nonlinear plastic hinges are made in beams at an offset from the panel zone-beam device interface, while column plastic hinges are associated at the panel zone-beam element interface. To account for P-delta effects, at each story stage, a leaning column is attached to the model via elastic beam-column elements and to the model via an axially rigid truss part. Rayleigh damping is assumed for a damping ratio of 2% during the first and third modes.

    The reference spectra for the MCE and DBE levels are generated utilizing the site's spectral acceleration values. As seen in Table 1, a total of seven ground motions were chosen from the PEER NGA database [65] and scaled according to ASCE 7-10. The earthquake records are scaled so that the average response spectrum for the chosen records is at least equal to the goal response spectrum for periods varying from 0.2 to 1.5 times the fundamental duration of the house. The nine-story building's steel members are chosen to meet the strength specifications of ANSI/AISC 360–10 [56] for the load combinations specified in ASCE 7-10. Additionally, the building is planned to meet ASCE 7-10 drift specifications. Due to the use of a nonlinear reaction experience in the architecture, the permitted story drift is raised by 25% and is calculated using ASCE 7-10 for Risk Category II buildings at 2.5 percent below the DBE level and 3.75 percent below the MCE level. Table 2 illustrates the column and beam pieces that were chosen for the nine-story structure. As shown below, the building meets the drift criteria for both DBE and MCE standards. The first mode cycle of the six-story and nine-story unique moment resisting frames is 2.04 seconds and 2.37 seconds, respectively.

**Table 1** Seven field motion recordings incorporated into the design.

| No. | Earthquake | Station name | Magnitude ($M_w$) | Distance (km) | Peak ground acceleration (g) |
|---|---|---|---|---|---|
| 1 | Northridge (1994) | Canyon Country | 6.7 | 12.4 | 0.40 |
| 2 | Kocaeli, Turkey (1999) | Bolu | 7.5 | 15.4 | 0.36 |
| 3 | Superstition Hills-02 (1987) | El Centro Imp. | 6.5 | 18.2 | 0.26 |
| 4 | San Fernando (1971) | LA-Hollywood | 6.6 | 22.8 | 0.19 |
| 5 | Duzce, Turkey (1999) | Duzce | 7.1 | 12.0 | 0.81 |
| 6 | Loma Prieta (1989) | Gilroy Array | 6.9 | 12.2 | 0.37 |
| 7 | Imperial valley-06 (1979) | El Centro array | 6.5 | 12.6 | 0.37 |

**Table 2** Members of nine story steel moment resisting frames

| Story | SMRF | | SMRF with SMARD | |
|---|---|---|---|---|
| | Exterior/Interior Columns | Girders | Exterior/Interior Columns | Girders |
| 1 | W18x311 | W21x201 | W18x311 | W21x201 |
| 2 | W18x311 | W21x201 | W18x311 | W21x201 |
| 3 | W18x311 | W21x201 | W18x283 | W21x182 |
| 4 | W18x311 | W21x201 | W18x283 | W21x182 |
| 5 | W18x283 | W21x182 | W18x234 | W18x192 |
| 6 | W18x283 | W21x182 | W18x234 | W18x192 |
| 7 | W18x234 | W18x192 | W18x192 | W18x143 |
| 8 | W18x234 | W18x192 | W18x192 | W18x143 |
| 9 | W18x192 | W18x175 | W18x143 | W18x130 |
| R | W18x192 | W18x175 | W18x143 | W18x130 |

### 7.2. Steel moment frames with SMARDs

To build the six- and nine-story frames with damping devices, a reduced-strength variant of the completely code-compliant structure is first produced. The column and beam component sizes are limited to the point that the steel frame meets the design code's strength specifications but does not exceed the drift restrictions. Additional dampers will hold a greater part of the dynamic excitation and will be primarily responsible for controlling tale drifts in this more compact structure. Table 2 lists the member sizes that were chosen for such a reduced strength frame. The reduced strength steel frame is initially updated with SMARDs to meet ASCE 7-10's tale drift specifications. The steel frame with SMARDs is constructed specifically utilizing the nonlinear reaction background protocol and in accordance with ASCE 7-10 Chapter 18 seismic architecture criteria for buildings with attenuating systems.

To model hysteretic response of NiTiHfPd SMAs at different temperatures and friction springs, a mechanical model is proposed in Open System for Earthquake Engineering Simulation (OpenSees) [59]. The model consists of two self-centering materials placed in parallel for the SMAs, as shown in Figure 13 (a) is proposed. The use of two self-centering elements instead of one for the SMAs provides more accurate modeling of the material response, especially for inner loops. Due to the dynamic loading that the SMAs would experience during an earthquake, the experimental responses of SMAs at 1 Hz are used for the numerical modeling. The selected parameters of the proposed model under the three temperatures of 25°C, 5°C, and -35°C are given in Table 3. Figure 13 (b) shows the hysteretic curves predicted by the proposed model and experimental response at the temperature of 25°C for the SMAS. Figure 13 (c-d) also shows the proposed model and experimental response for friction spring. The model consists of one self-centering material for friction spring. As can be shown, two proposed models can predict the response of NiTiHfPd SMA Material very well.

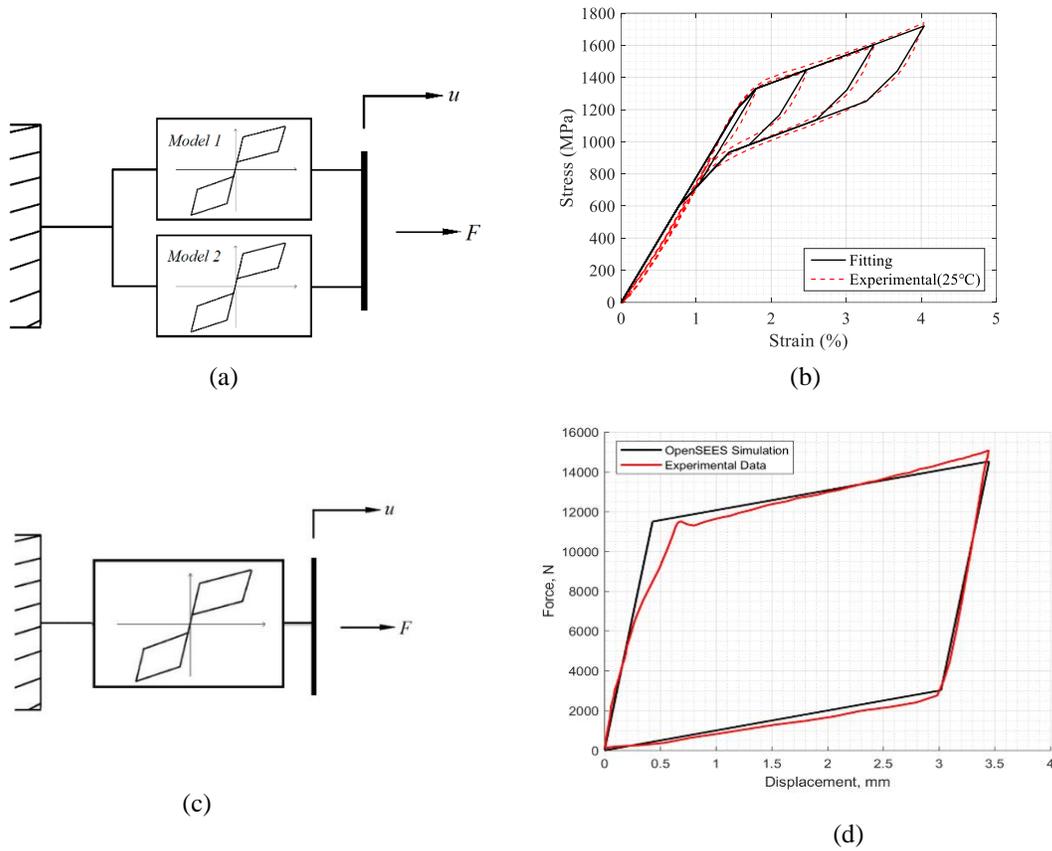

**Figure 13.** (a) A schematic representation for SMA bar model and (b) experimental stress-strain curve and model prediction at the temperature of 25°C for SMA (c) A schematic representation for friction spring model and (d) experimental force-deformation curve and model prediction for friction damper

Table 3 Parameters of proposed SMA model

| Temperature | 25°C | | 5°C | | -35°C | |
|---|---|---|---|---|---|---|
| Model | 1 | 2 | 1 | 2 | 1 | 2 |
| Initial stiffness (MPa/%) | 370 | 410 | 234.4 | 377 | 370 | 230 |
| Post-transformation stiffness (MPa/%) | 87 | 87 | 67.4 | 67.4 | 50 | 50 |
| Forward transformation stress (MPa) | 570 | 738 | 410.2 | 829.4 | 629 | 505 |
| Ratio of forward to reverse activation stress ($\beta$) | 0.50 | 0.20 | 0.70 | 0.29 | 0.80 | 0.40 |

The form memory alloy-based re-centering damper model is created in OpenSees by integrating multiple finite length elements with uniaxial self-centering material properties to reflect the SMAs and one finite length component with uniaxial self-centering material properties to represent the friction spring compound. The self-centering substance exhibits a flag-shaped hysteric reaction and exhibits post-transformation hardening activity with an initial stiffness similar to the post-hardening stiffness. The experimental findings for SMAs and friction spring compounds are used to establish model parameters (see Fig. 13). The following parameters were chosen for the self-centering content model of the SMAs: Original stiffness k1=61.23 kN/mm, post-activation stiffness k2=13.80 kN/mm, the ratio of post-transformation stiffening stiffness to initial stiffness =1.0, forward activating force Fa=104.01 kN, and forward to reverse activation force ratio=0.35. The following parameters were chosen for the friction springs' self-centering material model: Original stiffness k1 = 28.75 kN/mm, post-activation hardness k2 = 0.9677 kN/mm, ratio of post-transformation hardening stiffness to initial stiffness =1.0, forward activation force Fa=11.50 kN, and forward/reverse activation force ratio =1.0.

The SMARDs are located in two bays through each story level of the six-story house. The SMARDs are designed using the parameters defined in Section 6, and these characteristics are fixed for each unit. Each bay of a particular level is believed to have the same amount of dampers. According to figure 14, SMARDs are mounted utilizing a chevron brace design at the second and fourth bays of each story stage. Due to the high force potential of the SMARD system, the amount of dampers for each story is chosen to be two to meet the drift requirement. The average reaction of the structure with mounted dampers to the chosen seven ground motions meets the ASCE SEI 7-10 member intensity criterion and 2% drift specifications. The fundamental time of the six-story steel frame is 1.29 seconds, with the damper attached.

SMARDs are located at the 2nd and 3rd bays of each floor level of the nine-story house, figure 15, utilizing a chevron brace design as well. The amount of dampers for each tale is chosen to be two based on the nonlinear time history simulations in order to satisfy the drift specifications. The fundamental time of the nine-story structure is 2.19 seconds with the SMARDs installed.

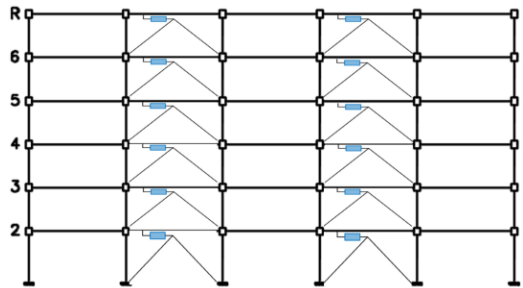

**Figure 14.** SMARD device installed into six-story steel frame

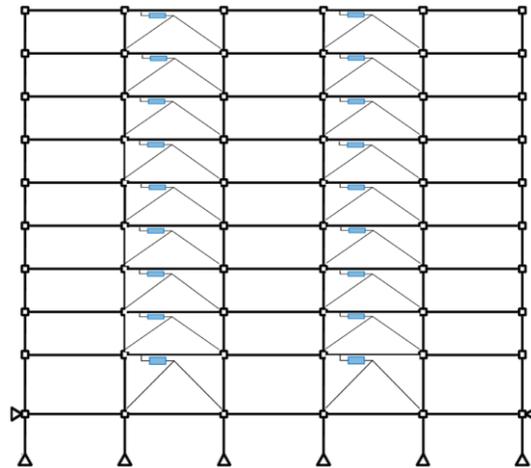

**Figure 15.** SMARD device installed into the nine-story steel frame

Table 4 shows the mean tale drifts for the SMRF and steel frame with SMARDs under seven concept ground movements at the DBE and MCE speeds. Notably, each frame is constructed in such a way that the peak inter-story drift acquired from the nonlinear response background technique satisfies the code drift specifications without striving for a higher level of seismic efficiency.

**Table 4** Drift response of developed frames under design earthquake records

| Model | $T_1$ (s) | Mean inter-story drift | | | |
|---|---|---|---|---|---|
| | | DBE | | MCE | |
| | | Design (%) | Limit (%) | Limit (%) | Limit |
| **SMRF (6-story)** | 2.04 | 2.38 | 2.5 | 3.21 | 3.75 |
| **SMRF (9-story)** | 2.37 | 2.44 | | 3.25 | |
| **6-story SMARDs** | 1.29 | 2.34 | | 3.19 | |
| **9-story SMARDs** | 2.19 | 2.39 | | 3.51 | |

## 8. PERFORMANCE ASSESSMENT OF THE BUILDINGS

In this analysis, the FEMA P695 technique [66] is used to conduct detailed nonlinear time history studies on both buildings using a series of 22 far-field ground motion pairs. The package contains high ground motions of PGA N 0.2 g and PGV N 15 cm/s, with event magnitudes ranging from M6.5 to M7.6 and originating in stiff soils (site Class D) and also very stiff soils (site Class C). Ground movements are standardized to exclude unjustified variation between documents. The earthquake ground records are scaled cumulatively to a predetermined danger point, as defined in ASCE/SEI 7-05 so that the median spectral acceleration of the record set matches the design spectra at the fundamental period of each six and nine-story frame (i.e., at 2.04 s for the unregulated six-story frame and 1.29 s for the regulated six-story frame). Additionally, at 2.37 seconds for unregulated nine-story frames and 2.19 seconds for managed nine-story frames). Peak inter-story drift ratio, which is associated with structural aspect damage; peak absolute floor acceleration, which is associated with non-structural portion damage; and peak residual story drift ratio, which again is associated with the structure's post-earthquake functionality, are chosen as the optimal reaction amounts. The residual tale drifts are determined by continuing the measurements for at least 20 seconds after the seismic incident has ended. As previous probability-based studies indicated, in computational simulations, a 10% inter-story drift ratio is used to describe the collapse [61].

### 8.1. six-story building

Under each ground motion record, building reactions to the unregulated structure and the building with mounted SMAs-based dampers are computed at two seismic hazard levels. The max inter-story drift ratio, spike residual inter-story drift ratio, and peak floor absolute acceleration for unregulated and managed buildings under DBE and MCE level ground motion records, respectively, are depicted in Figs. 16 and 17. As can be shown, the mounted SMARDs significantly minimize the drift requirement of a six-story building for all DBE levels of an earthquake. The collapse of an unregulated system is witnessed through nine ground motion records at the DBE stage.

Except in one situation when the peak inter-story drift ratio is 5.25 percent, the monitored structure's peak inter-story drift ratio is decreased below 3.25 percent at the same earthquake records. In eighteen earthquake events, a residual story drift of more than 1% existed in unregulated buildings. There are nearly no residual drifts found for the monitored building in all of the ground motion data, with the exception of earthquake number 24, which has a maximum residual drift of 1.24 percent. Nonetheless, in the majority of situations, the monitored building's peak story accelerations were slightly enhanced.

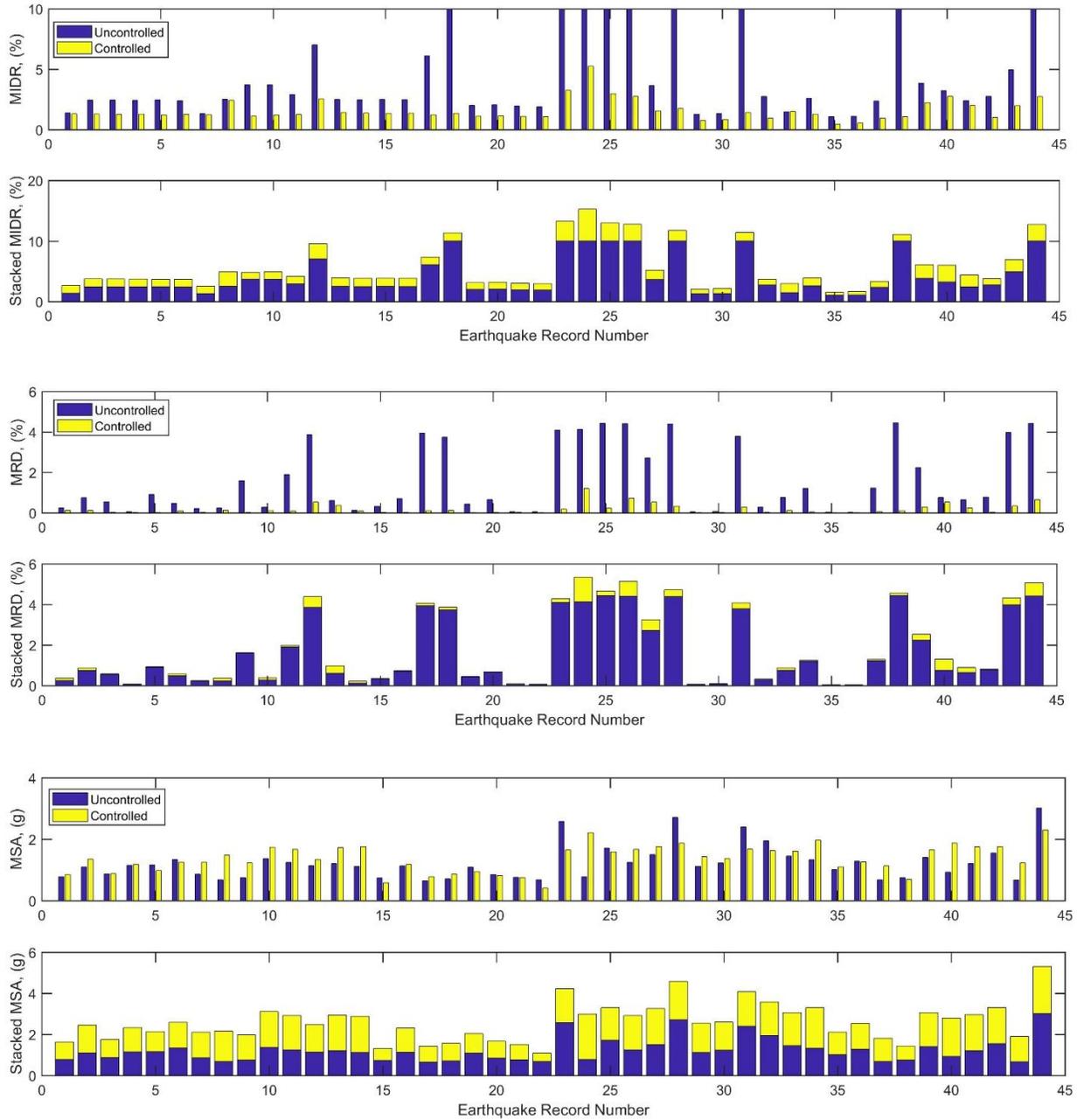

**Figure 16.** Max inter-story drift, Max residual story drift, and Max floor acceleration for individual DBE level ground motions.

Also, for MCE threshold seismic danger, the quantity of instances of structure failure increases sharply and reaches seventeen for unregulated buildings, despite the fact that the managed building collapses only once. It is noteworthy that the unregulated building exhibits significant permanent drifts in twenty-eight ground motion events. On the other side, in the majority of situations, a house revamped with SMARDs has very little residual drift. As with the DBE level danger, the monitored structure exhibits a greater acceleration response in over half of the ground shaking data scenarios.

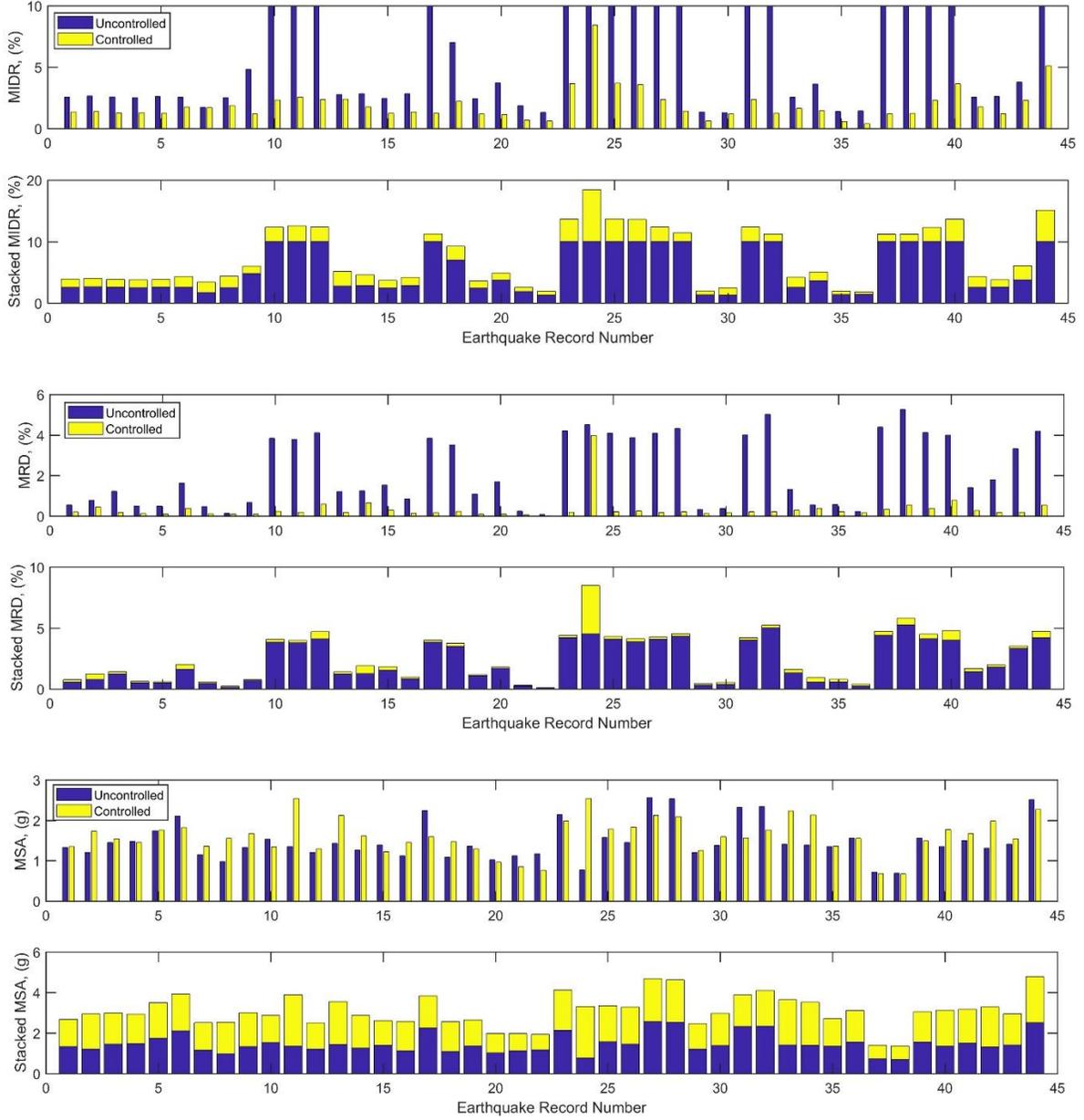

**Figure 17.** Peak inter-story drift, max residual story drift, and max floor acceleration for individual MCE level ground motions.

Due to the generally accepted lognormal distribution of seismic response, the figures for the peak floor absolute acceleration and peak inter-story drift ratio reaction for 44 earthquake reports are expressed in the form of the median (50th percentile) and 84th percentile responses and are as follows:

$$\widehat{X} = \exp\left(\frac{\sum_{i=1}^{n} \ln x_i}{n}\right) \quad (3)$$

$$X^{84} = \widehat{X} \exp(\sigma_{\ln X}) \quad (4)$$

Where $\sigma_{lnX}$ denotes the standard deviation of the logarithm of response X, and n represents the number of response data points. The 84$^{th}$ and median percentile values for the peak story absolute acceleration and peak inter-story drift ratio at each floor level for controlled and uncontrolled buildings up to the MCE and DBE levels are shown in Figure 18.

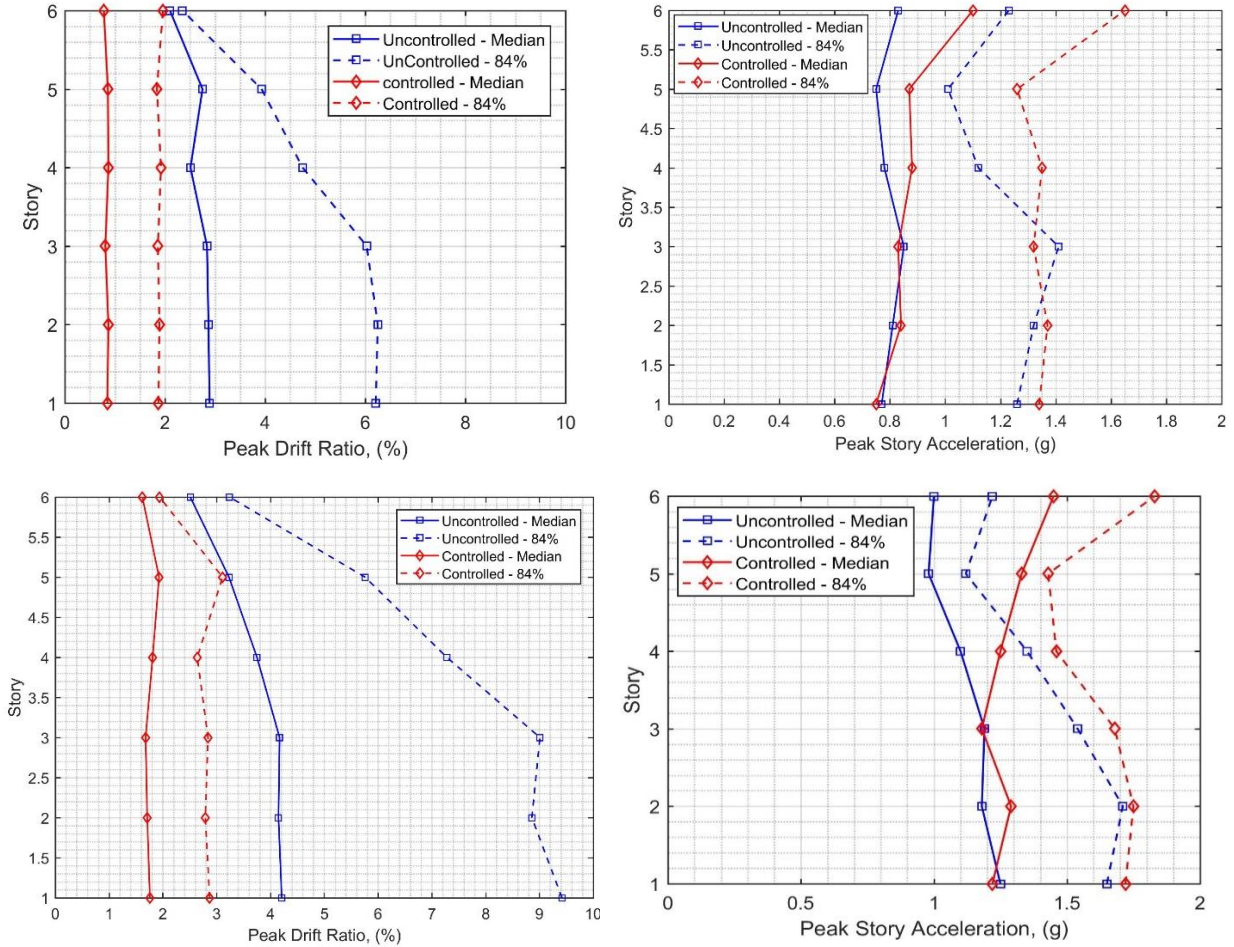

**Figure 18.** Statistics of max story absolute acceleration and max inter-story drift ratio for unprotected and controlled structures exposed to 44 earthquakes at the (a) DBE and (b) MCE levels

It can be shown that the peak inter-story drift values for the monitored system are lower for both stories for MCE and DBE danger levels. The maximum median peak inter-story drift ratio hits approximately 2.89 percent and 4.21 percent at the MCE and DBE ratios, consequently, for the unregulated framework. The median peak inter-story drift ratio, on the other hand, is evenly spread over the structure's height for the controlled frame, reaching a limit of 0.87 percent and 1.93 percent at the DBE and MCE stages, respectively. This demonstrates the efficacy of SMARDs in mitigating the displacement behavior of steel structures, especially at the higher seismic threat levels. Additionally, the maximum 84th percentile inter-story drift ratios for the traditional frame are 6.25 percent for the DBE and 9.42 percent for the MCE. The corresponding values for the steel frame with SMARDs are just 1.92 percent and 3.11 percent.

With the SMARD dampers attached, the steel frame's peak acceleration reaction is slightly increased. The DBE and MCE have peak median accelerations of 0.85 g and 1.25 g, respectively, and maximum 84-percentile acceleration values of 1.32 g and 1.71 g. The median acceleration values for the monitored frame are 0.88 g and 1.45 g at DBE, respectively, whereas the peak 84-percentile acceleration values are 1.65 g and 1.83 g at MCE.

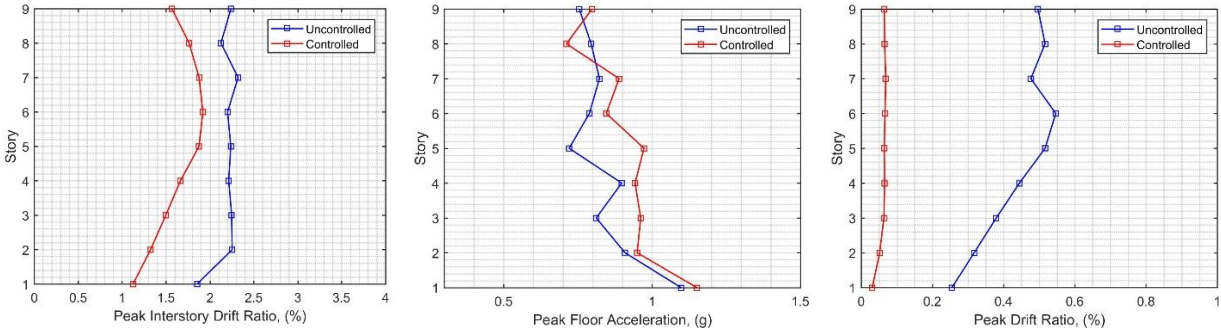
**Figure 19.** Envelopes for a median of response magnitudes for SMRF and SMARD frames at MCE level.

**8.2. Nine-story building**

Again, structural reactions of the unregulated structures and the structures with SMAs-based dampers are measured for the nine-story building from every ground motion at one seismic danger stage. To aid in comparing the efficiency of SMRF and SMARD schemes, Fig. 19 also includes the envelopes for such median peak response amounts under 44 ground-motion records for every device. As can be shown, each frame restricts the peak inter-story drift reaction satisfactorily under a variety of MCE level ground acceleration. The SMRF frame usually exhibits greater heterogeneity in systemic reaction from record to record, while the usage of SMARDs decreases response scatter. The peak median responses for the SMRF and SMARD frames are 2.32 percent and 1.92 percent, respectively. The maximum absolute acceleration responses at the top of the floor are 1.09 g and 1.15 g for both the SMRF and SMARD frames, respectively. Additionally, the SMARD frame exhibits modest residual drifts in the 9-story frame's height, while the SMRF device exhibits significant residual deformations.

## 9. CONCLUSIONS

The NiTiHfPd alloy and friction spring are included in this article as two critical components of a structural passive control device to provide re-centering and energy-dissipating capability in two steel frame buildings. To begin, experimental results are addressed for NiTiHfPd alloys with ultra-high power, strong damping power, and a broad operating temperature range. The effect of varying temperature ranges on the mechanical behavior of SMA bars is investigated. Additionally, experimental results are addressed for friction springs with a high force and deformation power. This study proposes a superelastic memory alloy re-centering damper system that takes advantage of the intrinsic re-centering capacity of high-performance SMA bars and the high energy dissipation potential of the friction spring compound to achieve superior seismic performance. For the re-centering and energy dissipation portion of the hybrid damper, dubbed the superelastic memory alloy re-centering damper, a highly damped friction spring compound is considered. Experimental experiments are conducted to describe the mechanical reaction of the SMARD device's subcomponents, namely the friction spring compound and high-performance SMA bars. Numerical analyses are used to determine the performance and efficacy of the suggested damper in minimizing the reaction of steel frame buildings to DBE and MCE level seismic loads. A six-story and a nine-story special steel moment frame construction are designed as a traditional moment resisting frame with superelastic memory alloy re-centering dampers attached. A set of 44 solid earthquake records is normalized and scaled in accordance with the procedures outlined in FEMA P695. We perform nonlinear reaction background studies and analyze peak response quantities. The results indicate that by installing SMARDs, the inter-story residual drifts and drift demands of buildings exposed to both DBE and MCE level earthquakes may substantially decrease without significantly increasing maximum acceleration demand. However, additional research is essential to validate the proposed device's response experimentally and investigate the impact of temperature on the hybrid device's response.

## Acknowledgments

This material is based upon the work supported by the National Science Foundation under Grant Number CMMI-1538770. The author of the paper also thanks my colleagues from the Department of Material Science at the University of Kentucky, who provided experimental data that greatly assisted the research.